# Lifting a sessile oil drop with an impacting one

Olinka Ramírez-Soto,[1,2,3,†] Vatsal Sanjay,[2,†] Detlef Lohse,[2,3] Jonathan T. Pham,[4*] and Doris Vollmer[1*]

1) Max Planck Institute for Polymer Research, Mainz, Germany.

2) Physics of Fluids Group, Max Planck Center for Complex Fluid Dynamics, Mesa+ Institute, and J.M. Burgers Center for Fluid Dynamics, University of Twente, Enschede, The Netherlands.

3) Max Planck Institute for Dynamics and Self-Organization, 37077 Göttingen, Germany.

4) Department of Chemical and Materials Engineering, University of Kentucky, Lexington, KY 40506, USA.

†) These two authors have contributed equally to this work

*) Corresponding authors: Doris Vollmer (vollmerd@mpip-mainz.mpg.de) and Jonathan Pham (Jonathan.Pham@uky.edu)

**Abstract**

Colliding drops are widely encountered in everyday technologies and natural processes, from combustion engines and commodity sprays to raindrops and cloud formation. The outcome of a collision depends on many factors, including the impact velocity and the degree of head-on alignment, in addition to intrinsic properties like surface tension. Yet little is known on the binary impact dynamics of low surface tension oil drops on a low-wetting surface. We experimentally and numerically investigate the dynamics of an oil drop impacting an identical sessile drop sitting on a superamphiphobic surface. We observe five rebound scenarios, four of which do not involve coalescence. We describe two previously unexplored cases for sessile oil drop lift-off, resulting from a drop-on-drop impact event. The simulations quantitatively reproduce all rebound scenarios and enable quantification of the velocity profiles, the energy transfer, and the viscous dissipation. Our results illustrate how varying the relative offset and the impact velocity results in controllable rebound dynamics for low surface tension drop collisions on superamphiphobic surfaces.

**One Sentence Summary:**

We experimentally and numerically determine and quantitatively model when and how an impacting oil drop lifts a sessile drop, consisting of the same oil, resting on a superamphiphobic surface.

## Introduction

When a liquid drop impacts a sessile one of an identical liquid, it is intuitively expected that both drops coalesce. This process is commonly observed in day-to-day examples, such as rain or drops from a leaky faucet. However, coalescence can be obstructed by a thin layer of air between the two drops (*1–3*). Insufficient thinning of this air layer during impact even enables water drops to bounce from perfectly hydrophilic surfaces, which they would otherwise wet (*4–6*). In the late 1800s, Reynolds (*7*) noticed that water drops can glide over a pool because of this air layer. A vapor layer also governs the Leidenfrost effect (*8–10*), where a drop hovers over a superheated surface. As a result, drop bouncing, coalescence, and spreading can all be observed depending on the intrinsic properties of the liquid, as well as external parameters, such as the background pressure, collision velocity, and the relative impact parameter describing whether the collision is head-on or off-centered (*11*, *12*, *21–23*, *13–20*). Despite this progress in the experimental characterization of the impact dynamics, a quantitative modelling of the velocity fields and energy transfer is lacking, especially for non-aqueous liquids.

Drop impact on surfaces, and the outcome of the collision, is of practical importance for many situations. For example, in agriculture, it is essential to ensure that pesticides and other chemicals sprayed on wet leaves do not roll off and contaminate the surroundings (*24*). On the other hand, removal of drops is desirable for car windows (*25*) and self-cleaning of surfaces. On superhydrophobic surfaces, a water drop impacting another one can lead to drop removal after coalescence. Sufficient transfer of kinetic energy from the impact event turns the two drops into a single merged drop and leads to bouncing after coalescence (*22*, *26–30*). Alternatively, both drops can also rebound from the surface if sufficient energy is exchanged during impact (*17*).

While several reports exist on understanding how a water drop impacts a sessile water drop on a surface (*15*, *17*, *31–33*), the dynamics of a low surface tension oil drop impacting an oil drop on a non-wetting surface remains unexplored. It has been shown that the collisional dynamics of free-flying oil drops offer more diverse outcomes than those of water drops (*34*). Does this also hold in the presence of a low-wetting surface? What resulting scenarios exist for drop-on-drop impact of oil on a superamphiphobic surface? How is energy transferred between the drops? Intuitively, the rebound of oil drops from a surface by impact with another oil drop seems unlikely for the following reasons. 1) The surface tension $\gamma$ of most hydrocarbon oils (25 mN/m) is significantly lower than that of water (72 mN/m). Smaller $\gamma$ reduces the transfer of surface energy to kinetic energy during the coalescence. This implies that the droplets have less energy to rebound. 2) Large sessile oil drops typically have a large contact size. On a flat surface, the receding contact angle is typically below 60° and often close to zero (*35*). Consequently, receding oil drops easily rupture before coming off the surface. 3) On a superamphiphobic surface, oil drops display large apparent contact angles (*36*, *37*). However, the true liquid-solid contact angle is still small, leaving oil drops in a metastable state. Pressure as low as a few hundred Pascal is sufficient to transition the drop from the metastable Cassie state to wet the surface thoroughly (*37*, *38*). 4) The low surface tension of oil means that the drop is easily deformable, which may give rise to enhanced viscous dissipation and energy loss upon impact.

In this contribution, we experimentally and numerically investigate the dynamics of a low surface tension oil drop impacting a sessile drop of the same liquid, resting on a superamphiphobic surface (Fig. 1a). Indeed, we find that the impacting oil drop can lift the resting drop off the surface, without ever coalescing. Notably, we find four rebound scenarios without coalescence: (i) both drops rebound, (ii) two scenarios where the impacting drop rebounds while the sessile drop remains, and (iii) the sessile drop rebounds while the impacting drop remains on the surface. We illustrate how these impact outcomes are governed by the Weber number and the extent of offset from a head-on collision. Direct numerical simulations provide a quantitative description of the velocity fields in both drops and how energy is transferred between the two drops during impact.

## Results

### Method

In our experiments, a sessile oil drop is gently positioned on a superamphiphobic surface and then impacted with a second identical oil drop (Fig. 1a). The superamphiphobic surface is composed of a ~20 μm thick layer of templated candle-soot (*10*, *39*). Candle soot consists of a porous network of 50±20 nm sized carbon nanobeads. Making use of chemical vapor deposition (CVD) of tetraethyl orthosilicate (TEOS) catalyzed by ammonia, a ~25 nm thick layer of silica is deposited over the porous nanostructures to increase the mechanical stability of the fragile network (Fig. 1a-i, fig. S1). The soot-templated silica network is fluorinated with trichloroperfluoroctylsilane to lower the surface energy, producing a superamphiphobic surface which repels water and most oils. A drop of hexadecane (Fig. 1a-ii) exhibits an apparent contact angle of $\Theta^{app} = 164° \pm 1°$, an apparent receding contact angle of $\Theta_r^{app} = 158° \pm 3°$, and an apparent advancing contact angle of $\Theta_a^{app} \approx 180°$ (*40*), as determined by confocal microscopy (Fig. 1a-iii and fig. S2, fig. S3). Low lateral adhesion of hexadecane is confirmed by measuring a low roll-off angle of $\alpha = 3° \pm 2°$ (*41*).

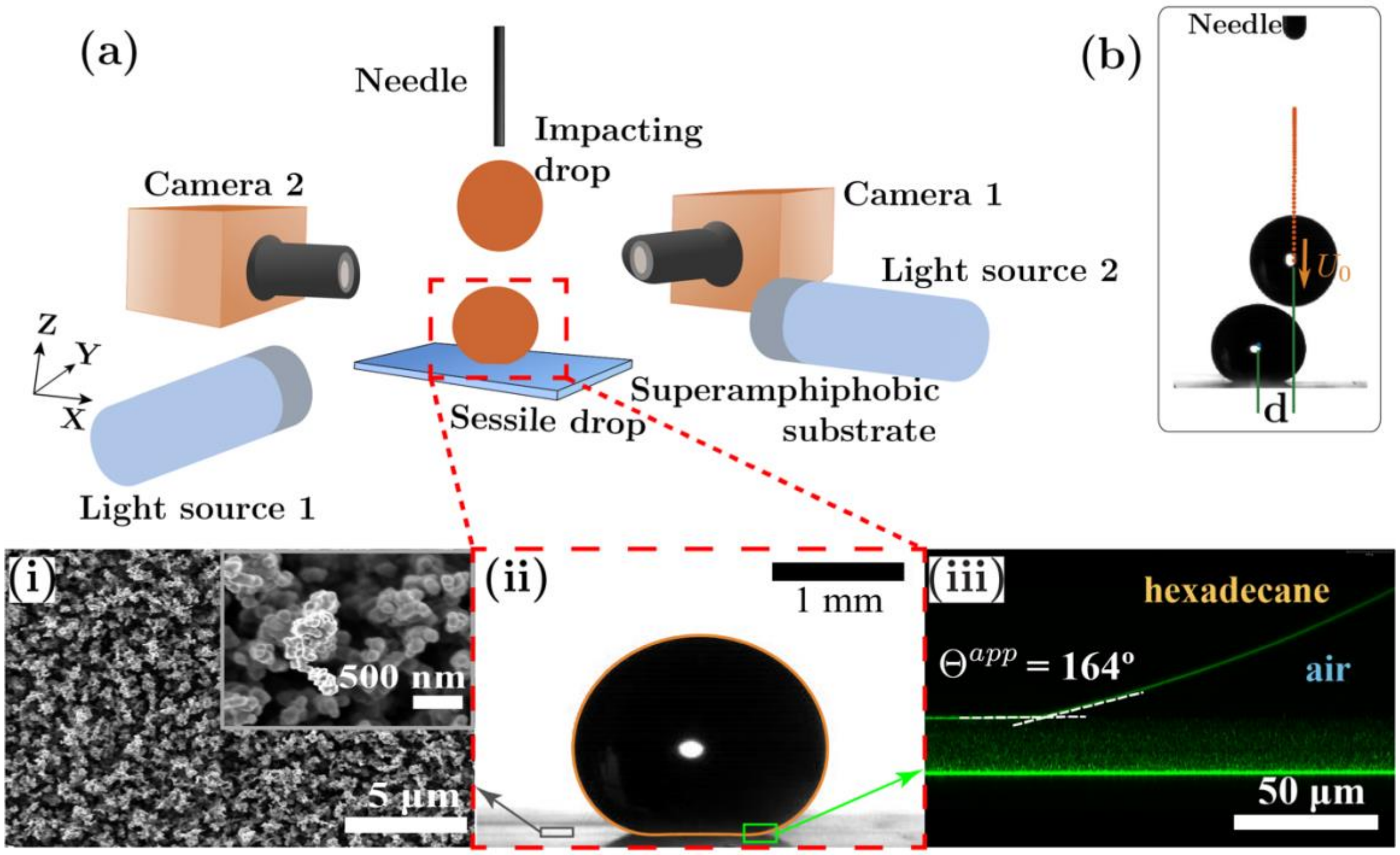

**Fig. 1. Experimental approach and the sessile drop: (a)** Sketch of the experimental setup for binary drop impact on superamphiphobic surfaces. The needle is fixed to set the impacting height in the Z direction, and the relative distance between the sessile and impacting drops. The sessile drop is first centered along the YZ plane. Then the impacting drop is dispensed from the needle while the impact is monitored with camera 2. Camera 1 is used to determine the relative positions in the X direction. The cameras and the light sources are aligned to observe the impact both in the XZ and YZ planes. **Insets: (i)** SEM

image of a soot-templated surface at two magnifications. **(ii)** Hexadecane drop ($V \approx 3$ µL) resting on the superamphiphobic surface. The orange contour is the solution of Eq. 1 for a corresponding Bond number $Bo = 0.3$. **(iii)** An inverted laser scanning confocal microscope image illustrating the apparent contact angle. Reflection of the interfaces are shown in green. **(b)** Image showing an off-center collision. The impact parameter is $\chi = d/(2R)$.

For our drop impact studies, a sessile drop of hexadecane is gently placed on this superamphiphobic surface with a needle connected to a syringe pump (dosing rate: 2 mL/h). When gravity exceeds the drop-needle adhesion, the drop releases from the needle; this results in a drop volume of $V \approx 3$ µL (Fig. 1a). This volume corresponds to a Bond number of 0.3 ($Bo = \rho_l g R^2/\gamma$, where $\rho_l$ is the density of the liquid, $g$ is the gravitational acceleration and $R$ is the radius of a spherical droplet of identical volume). The Bond number relates inertia to surface energy, reflecting how gravity affects the shape of the sessile drop. This shape is important as it forms the initial condition for the numerical simulation. To calculate and confirm this shape numerically, we solved the Young – Laplace equation.

$$-\frac{\partial P'}{\partial X_i} + (\kappa - \Delta\rho\, Bo\, Z)\delta_s n_i = 0 \qquad (1)$$

In Eq. 1, $P'$ refers to the reduced pressure (as defined in (*42*)), $X_j$ refers to the coordinate system unit vector, $Bo$ Z is the gravitational potential, $\kappa$ the curvature of the liquid-gas interface, $\Delta\rho$ the normalized density difference across this interface (non-dimensionalized with $\rho_l$), $\delta_s$ the Kronecker delta function (1 at the interface and 0 otherwise), and $n_i$ the unit vector normal to the interface. Note that all equations in this manuscript are written using the cartesian tensor notation. The shape of the drop is calculated by solving Eq. 1 and matches well with experiments (Fig. 1a-ii).

The control parameters of the drop collision, determining the outcome, are the Weber number ($We$), which is related to the impact velocity ($U_0$), and the impact parameter ($\chi$), which describes the relative offset position of the two colliding drops. The impact velocity $U_0$ is controlled by positioning the needle to a defined height (Fig. 1a). The corresponding Weber number $We = \rho_l U_0^2 R^2/\gamma$ compares inertia and surface tension, where $\rho_l = 770$ kg/m3 is the density of the hexadecane and $\gamma = 27.5$ mN/m is the surface tension. In our experiments, the Weber number ranges from 0.02 to 9. The substrate is then translated laterally to position the drop in the X and Y directions. At an identical dosing rate, a second drop is released with an identical volume, $V \approx 3$ µL, and impacts the sessile drop. Two high-speed cameras are perpendicularly positioned to capture the dynamics of the drops in the X, Y, and Z directions. The offset position of the two drops is given by the ratio ($\chi = d/(2R)$), where $d$ is the horizontal difference of the center of masses of the impacting drop and the sessile drop (Fig. 1b). $\chi = 0$ describes a perfect head-on collision whereas $\chi = 1$ corresponds to the situation when the two drops merely brush each-other ($d = 2R$).

**Experimental Observations:**

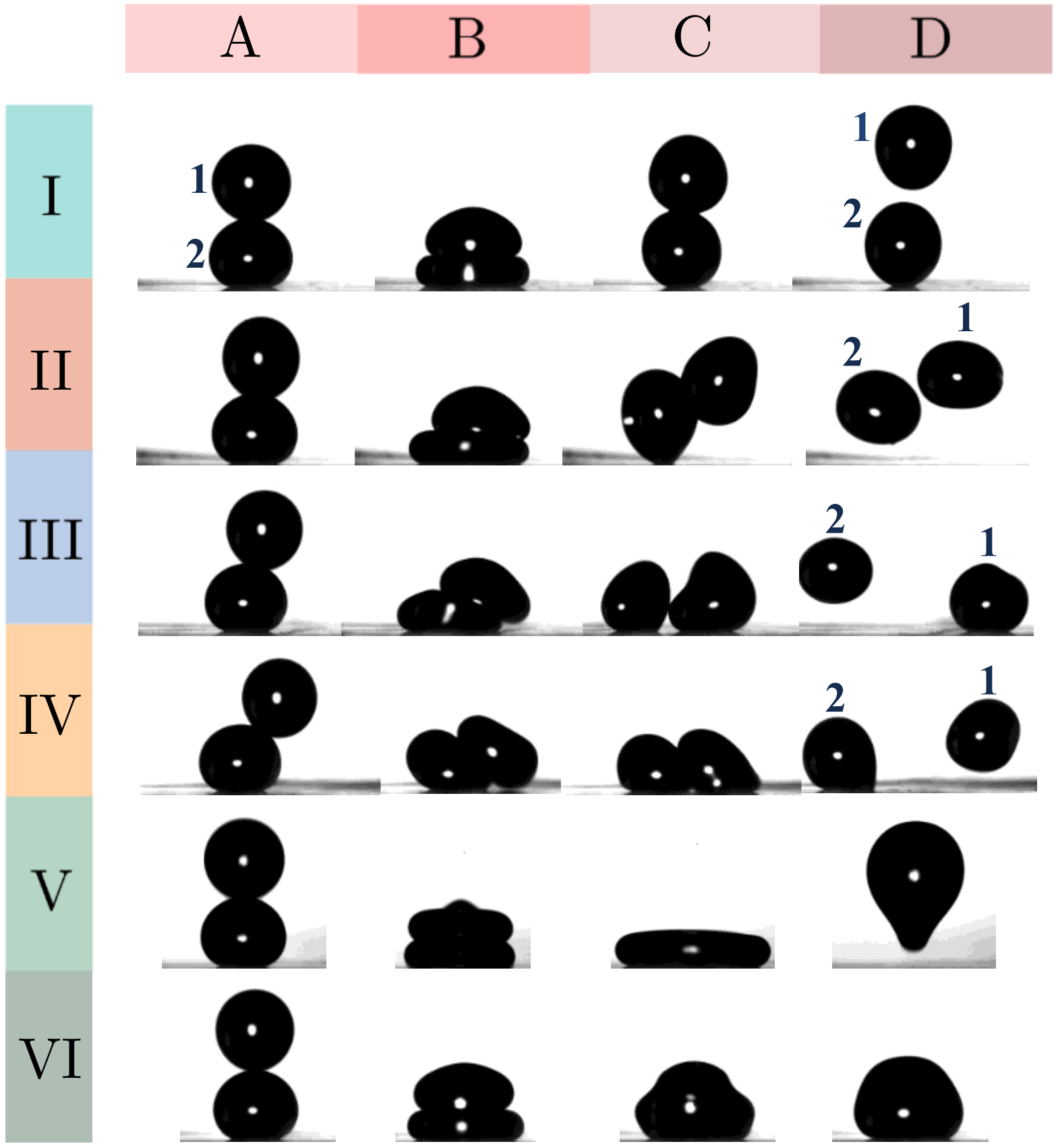

**Fig. 2. Snapshots of the impact dynamics:** Note that the drop labels 1 and 2 are for the impacting and sessile drop, respectively. Six outcomes (Cases I – VI) are observed when varying the offset $\chi$ and the Weber number ($We$). The rows correspond to different offset parameter for I-IV. The columns show characteristic stages of the collision process. **A:** just at collision, **B:** sessile drop at maximum compression, **C:** droplet shape just before separation or coalescence. **D:** final outcome of the impact. The height of the center of mass of the impacting, sessile, or coalesced drops are maximal. Volume of both drops is 3 μL. Case I, $We$ = 1.30 and $\chi$ = 0.01, the time stamp for each frame is: $t_A$ = 0 ms, $t_B$ = 8 ms, $t_C$ = 20 ms, $t_D$ = 25 ms. Case II, $We$ = 1.53, $\chi$ = 0.08. $t_A$ = 0 ms, $t_B$ = 8 ms, $t_C$ =20 ms, $t_D$ = 24 ms. Case III, $We$ = 1.44, $\chi$ = 0.24, $t_A$ = 0 ms, $t_B$ = 8 ms, $t_C$ = 20 ms, $t_D$ = 24 ms. Case IV, $We$ = 1.48, $\chi$ = 0.52, $t_A$ = 0 ms, $t_B$ = 5.5 ms, $t_C$ = 7 ms, $t_D$ = 21 ms. Case V, $We$ = 5.84, $\chi$ = 0.08, $t_A$ = 0 ms, $t_B$ = 3.75 ms, $t_C$ = 8.5 ms, $t_D$ = 25.5 ms. Case VI, $We$ = 1.43, $\chi$ = 0.03, $t_A$ = 0 ms, $t_B$ = 7.5 ms, $t_C$ = 9 ms, $t_D$ = 17 ms.

When varying the offset position $\chi$ and the Weber number $We$, six outcomes for the impact dynamics are observed, termed Cases I-VI (Fig. 2). The column A of images is taken just as the collision starts ($t = 0$ ms) and is used to quantify the offset position, $\chi$. Column B is at the point of maximum sessile drop compression, and column C demonstrates the shape of both drop just before they separate or coalesce. Column D illustrates the overall outcome of the collision event. We first consider the outcomes at $We \approx 1.5$ while varying $\chi$. For a near zero $\chi$, Case I is observed, which is a head-on collision (Fig. 2, Supplementary Movies 1-3, fig. S4). During impact, both drops deform and spread radially, and as a result, show axial compression. The kinetic energy of the system is transferred to the surface energies of both deformed drops. Moving forward in time, both drops start to retract. The sessile drop transfers energy back to the impacting drop in the form of kinetic energy. Upon completion of the collision, the impacting drop bounces off while the sessile drop stays on the substrate. The sessile drop oscillates, hinting that it retains a part of the energy gained during impact. For a slightly higher offset, $\chi \lesssim 0.15$, Case II is observed (Supplementary Movies 4-6, fig. S5). The initial collision is similar to Case I in that the drops collide, followed by vertical compression and lateral spreading. However, unlike Case I, the deformations are no longer symmetric, and the sessile drop also lifts off the surface. The displacement for either drop with respect to the center of mass of the initial sessile drop is in opposing lateral directions. Further increasing of the offset to $\chi \lesssim 0.5$, the impacting drop glides over the sessile drop and rolls on the substrate, as illustrated by Case III (Fig. 2, $\chi = 0.24$, Supplementary Movies 7-9, fig. S6). Unlike Cases I and II, no rebound of the impacting drop is observed. Surprisingly, instead the sessile drop lifts-off the surface. As the offset value is increased even further ($\chi > 0.5$, Case IV), the impacting drop still rolls over the sessile drop (Supplementary Movies 10-12, fig. S7). However, during retraction, the impacting drop rebounds from the surface while the sessile drop moves along the surface.

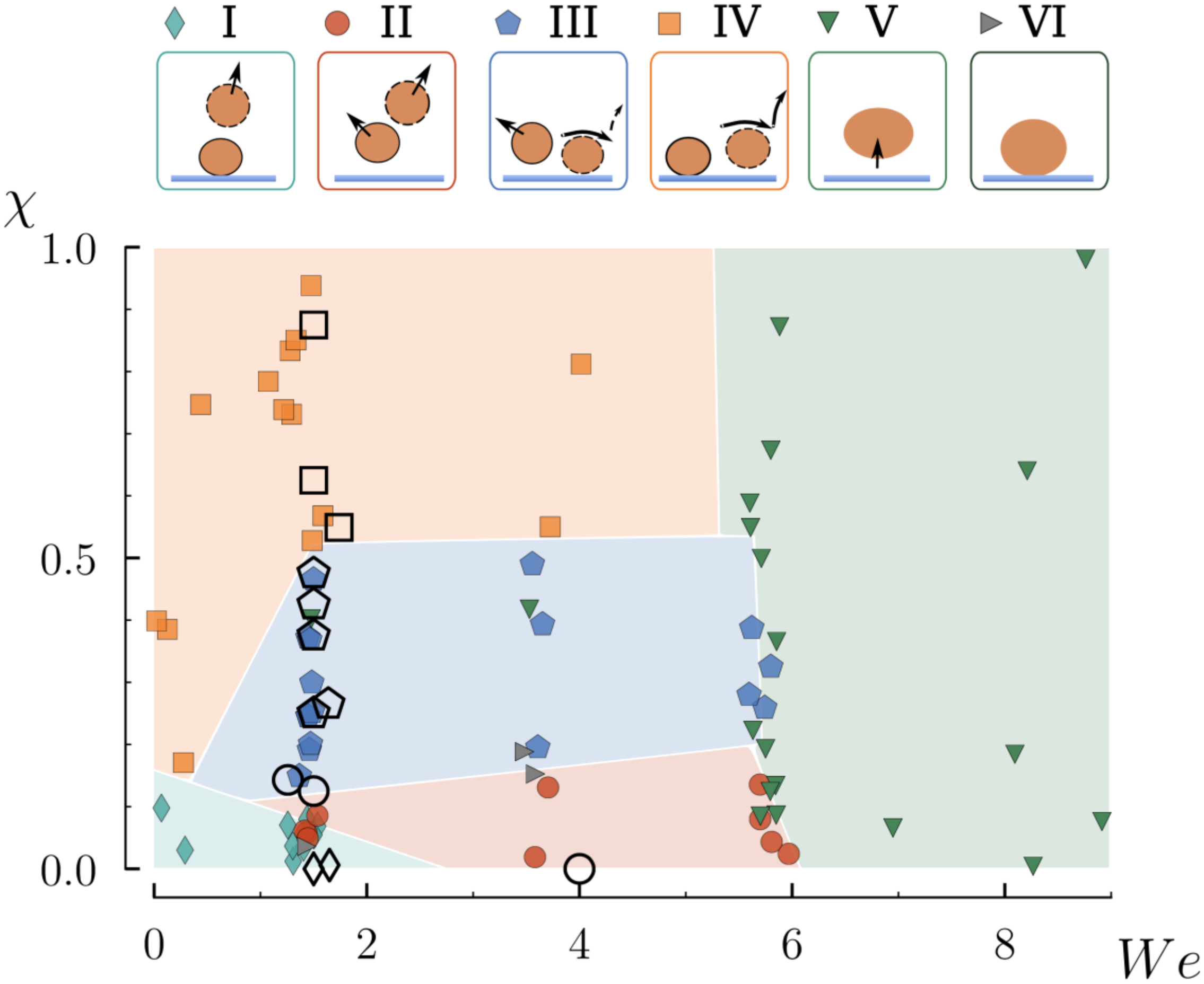

**Fig. 3. Regime Map:** Phase diagram illustrating the observed cases as a function of the offset parameter $\chi = d/(2R)$ and Weber number $We$. The top sketches with the respective Roman case number are the possible outcomes after the hexadecane drop impacted on the sessile hexadecane drop. Each possible outcome is marked by a color and symbol for identification and corresponds to the sketched cases I-VI. Closed symbols correspond to experiments and open ones to numerical simulations.

In the above Cases I-IV, the Weber numbers were kept constant at $We \sim 1.5$ while the offset was varied. However, the outcome of the impact event also varies with the Weber number. To provide a better intuition on how both $\chi$ and $We$ affect the observed outcomes, we plot our data as a phase diagram (Fig. 3). When the Weber number is increased above $We \geq 6$, regardless of the offset parameter $\chi$, we find coalescence of the two drops, as illustrated in Case V (Fig. 2, Supplementary Movie 13, fig. S8). In this regime, the air layer between the drops is unstable which results in direct contact and subsequent coalescence. The coalesced drop reaches a maximum spreading diameter during impact (column C in Fig. 2). During retraction, the drop elongates vertically and ultimately detaches from the surface. Occasionally, drops coalesce without subsequent bouncing (Case VI, Supplementary Movie 14, fig. S9). Although this outcome is rarely observed and likely caused by surface defects, we present this result for the sake of completeness to demonstrate all observed outcomes.

**Direct Numerical Simulations:**

Although the experimental observations consistently illustrate how $We$ and $\chi$ dictate the observed impact outcomes, they lack detailed information on the velocity fields and on how energy is transferred between both drops. To ascertain this information, we ran Direct Numerical Simulations (DNS) and compared these results with our experimental data.

We first ran four simulations choosing $We$ and $\chi$ values within the regimes for Cases I-IV, as denoted by open symbols in Fig. 3. The results are displayed in Fig. 4. The normalized times ($t^* = t/t_\gamma$, where $t_\gamma$ is the inertial-capillary time scale, $\sqrt{(\rho R^3)/\gamma}$) correspond to the stages of the process, as described by columns A – D in Fig. 2. As is evident from the top rows (orange drops), the simulations reproduce the general collision outcomes consistent with the snapshots of the impact dynamics (Fig. 3). Moreover, the direct numerical simulations allow for quantifying the velocity vector fields for each of the cases (Fig. 4, bottom rows). These vector fields, combined with a calculation of the energy budget, renders it possible to quantitatively explore the dynamics of the oil drop-on-drop collision process. To account for the kinetic energy $(E_k)$, gravitational potential energy $(E_p)$, surface energy $(E_s)$ and dissipative losses $(E_d)$, we numerically calculated the total energy of the system as

$$E = E_m + E_s + E_d \tag{2}$$

In Eq. 2, the total mechanical energy $E_m = E_k + E_p$, the surface energy $E_s$, and the energy dissipation $E_d$ are calculated using a method similar to the one developed by Wildeman et al. (*43*). $E_k$ includes the kinetic energy of the center of mass as well as the oscillation and rotational energies obtained in the reference frame that is translating with the center of mass of the individual drops. The details of these calculations are provided in the *Material and Methods* section.

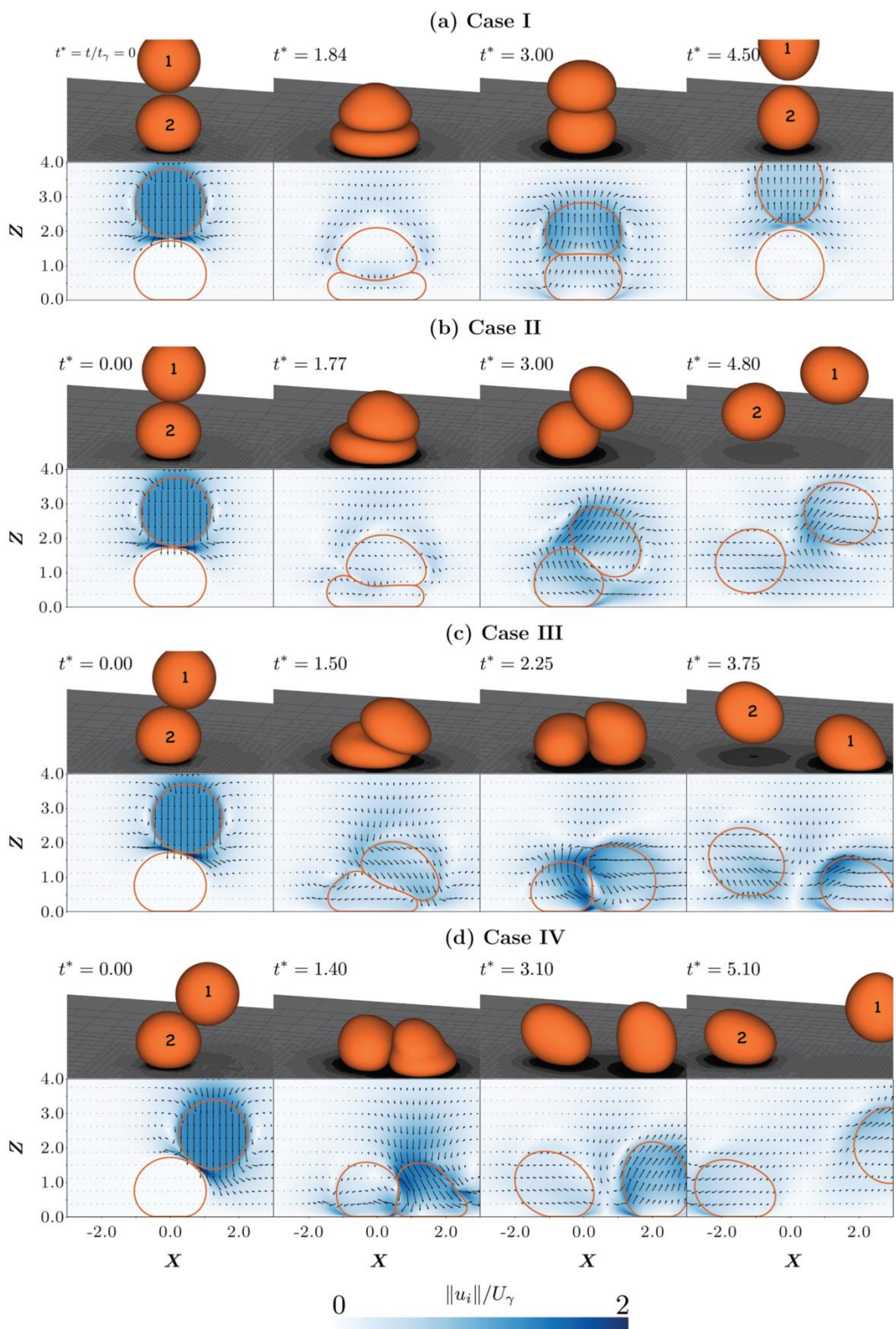

**Fig. 4. Snapshots of Direct Numerical Simulations:** Illustration of different phases of drop-on-drop collisions and the subsequent outcomes. **(a)** Case I: ($\chi = 0$) impacting drop

bounces back and the sessile drops stays on the substrate, **(b)** Case II: ($\chi = 0.08$) impacting drop bounces back and the sessile drop lifts-off from the substrate, **(c)** Case III: ($\chi = 0.25$) impacting drop stays on the substrate and the sessile drop lifts-off, and **(d)** Case IV: ($\chi = 0.625$) impacting drop bounces back and sessile drop stays on the substrate. For all these cases, $We = 1.5$. The drop labels 1 and 2 are for the impacting and sessile drop, respectively. t* is the non-dimensionalized time used for the numerical simulations and is given by $t = t/t_\gamma$ where $t_\gamma$ is the inertial-capillary time scale, $\sqrt{(\rho R^3)/\gamma}$. The absolute values of the normalized velocities vary between zero (white) and twice the inertial-capillary velocity, $U_\gamma = \sqrt{\gamma/(\rho R)}$ (dark blue).

While keeping the Weber number at $We \sim O(1)$, the cases appear in order from I to IV with increasing offset position $\chi$. For all cases, the energy is initially contained in the mechanical energy of the impacting drop (i.e. its kinetic and potential energy) and the surface energy of the sessile drop. To describe the system energy of the DNS results presented in Fig. 4, we plot the full energy balances for each case in Fig. 5. For comparison convenience, the energies in Fig. 5 are normalized with this initial energy of the system.

Let us consider first a head-on collision where $\chi = 0$ (Fig. 4a, Fig. 5a, and Supplementary Videos 2-3, Case I), which is defined by a symmetric configuration. First, the momentum is transferred from the impacting drop to the sessile drop, as the sessile drop deforms. This transfer results in deceleration of the impacting drop. Moreover, the kinetic energy of the impacting drop transforms into the surface energy of the system. This transfer continues until $t^* = 1.84$ (Fig. 4a: Column B) when the deformation in the two drops is maximum. Even at the moment of maximal elongation of both drops, the kinetic energy remains finite because of rotational flow within the drops (Fig. 4a: Column B, velocity field) (*43*). The mechanical energy passes a minimum ($t^* = 1.84$) when the surface energy is maximal. For $t^* > 1.84$, the surface energy of the two drops is converted back into kinetic energy. Retraction of the sessile drop is hindered by the impacting one (Fig. 4a: Column C), directly sitting on top of it. As a result, the sessile drop cannot lift-off from the substrate, but it releases any extra energy by oscillations (Supplementary Videos 1-2). During impact, the drops lose approximately 20% of their initial energy through viscous dissipation inside the drops and the thin air layer between them (Fig. 5a). This dissipation occurs mainly during the initial stages of the process ($t^* < 3$). It should be noted that the surface tension ($\gamma$), viscosity ($\mu$) and impact velocity ($U_0$) all affect viscous dissipation (*44*). These properties are related to the Ohnesorge number ($\mathrm{Oh} = \mu/\sqrt{\rho\gamma R} \approx 0.03$), which compares viscous and surface tension forces, and the Weber number, $\mathrm{We} = \rho U_0^2 R/\gamma \sim \mathcal{O}(1)$ (see Eq. 16 and (*45*)). The dissipation observed in our case is lower than that reported previously for a single drop impact at comparable Oh and We on superhydrophobic (*43*) and superamphiphobic substrates (*44*). In the case of a single drop impact, the velocity of the drop goes to zero quickly as it approaches a rigid substrate (*46*), leading to high dissipation close to the substrate (in the thin air layer and near the contact line). In the case of drop-on-drop impact, the sessile drop is deformable, decreasing the deceleration experienced by the impacting drop. As a result, the system retains almost 80% of its initial energy in the form of mechanical and surface energy of the drops.

For slightly off-center collisions where $\chi = 0.08$ (Fig. 4b, Fig 5b, and Supplementary Videos 5-6, Case II), the initial collision is similar to Case I; the drops collide, followed by vertical compression and lateral spreading. However, unlike Case I, the impacting and the sessile drops lift-off from the substrate. This feature results from the loss of axial symmetry of the velocity field for $\chi > 0$. During retraction, transfer of momentum from the compressed sessile drop back to the impacting drop occurs mainly along a vector pointing normal to the apparent contact zone. Moreover, the sessile drop attempts to regain its spherical shape (minimum surface energy state). As a result, the velocity field of the sessile drop is almost parallel to the contact zone, i.e. pointing to the upper left. These opposing orientations of the velocity fields cause the impacting drop to bounce off the sessile drop, and the sessile drop to lift-off from the substrate. (See the velocity vector fields in Fig. 4b and Supplementary Video 6). Viscous dissipation increases compared to a head-on-collision, but still is maximum during the initial stages of the process ($t^* < 3.5$, Fig. 5b).

As the offset is further increased to $\chi = 0.25$ (Fig. 4c, Fig. 5c, and Supplementary Videos 8-9, Case III), the impacting drop glides over the sessile drop (facilitated by the thin air layer), and sufficient energy is transferred to lift the sessile drop from the substrate. This can be understood from the interplay of the velocity field and the contact time (Fig. 4c and Supplementary Video 9). The relatively large offset causes the averaged velocity field of the restoring impacting drop to point both almost parallel to the surface and downwards, while the velocity field of the sessile drop is pointing upwards. The large deformations of both drops are reflected in the evolution of the surface energy (Fig. 5c). The large deformations of both drops also causes an increase in viscous dissipation ($E_d$); at the end of the process, almost 50% of the initial energy is lost. Moreover, unlike cases I and II, viscous dissipation not only occurs in the drops, but also in the thin air layer as the impacting drop approaches the substrate (*4*).

Finally, if the offset is increased even more to $\chi = 0.625$ (Fig. 4d, Fig. 5d, and Supplementary Videos 11-12, Case IV), the time of contact is insufficient to transfer enough energy to the sessile drop for lift-off (*47*). Moreover, the vector normal to the drop-drop contact area is farthest from vertical as compared to the normal vectors in other cases. That is, it points nearly horizontal. As a result, the sessile drop rolls along the substrate and the impacting drop instead rebounds from the surface, resembling typical drop-surface impact. In this case, most of the energy is retained by the impacting drop, as illustrated in Fig. 5d. Similar to Case III, viscous dissipation accounts for almost 50% of the initial total energy. Although in Case I and IV the impacting drop rebounds while the sessile drop remains on the surface, we discriminate between both cases. For Case I, the vector fields are symmetric around the X = Y = 0 axis, whereas for Case IV the vector fields are highly asymmetric and the sessile drop rolls along the surface. Furthermore, in Case IV, the impacting drop bounces-off the substrate, as opposed to the sessile drop in Case I.

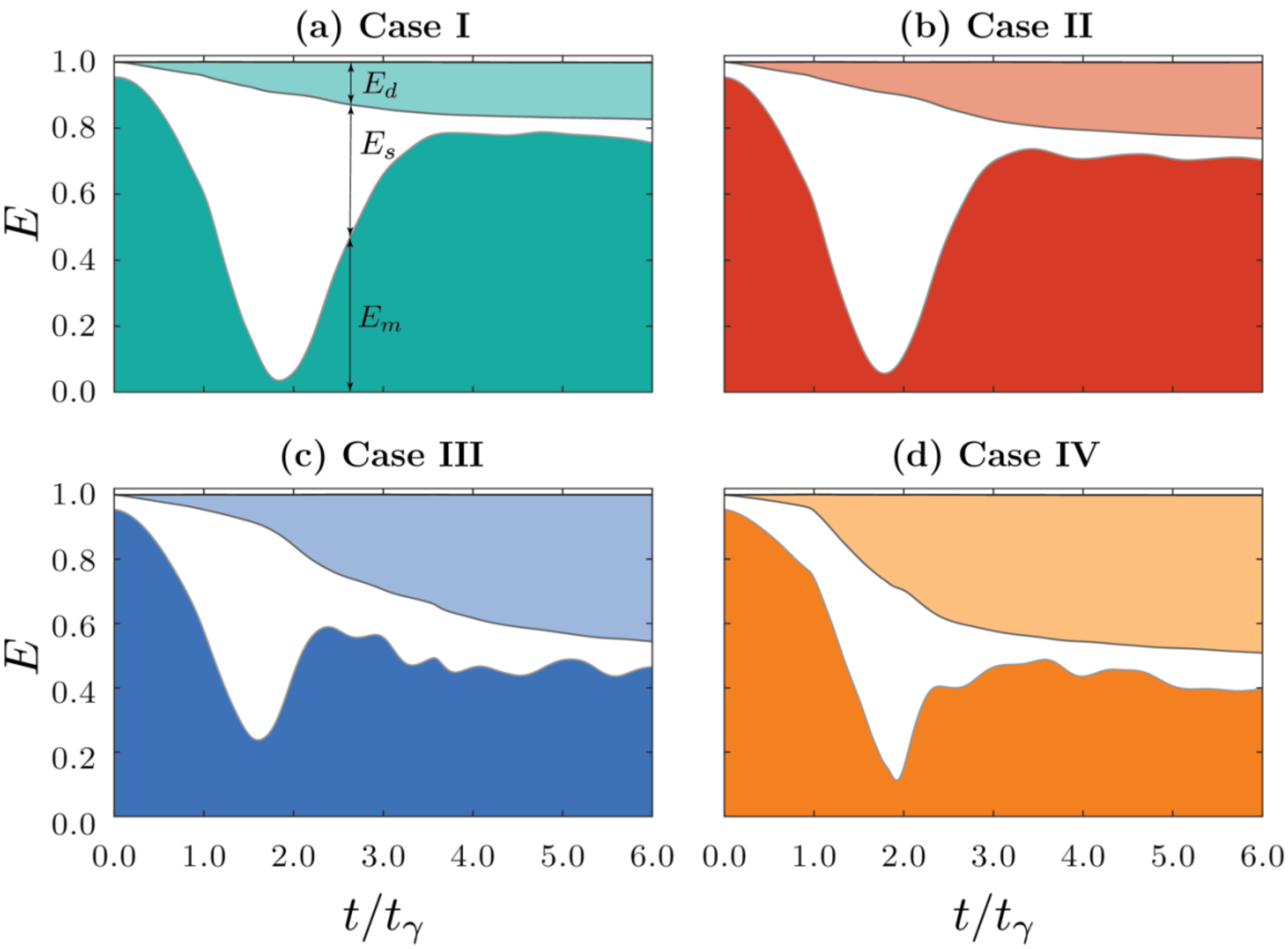

**Fig. 5. Energy Budget:** The temporal variation of energy transfer elucidates different stages of the drop-on-drop impact process at $We \sim 1$. Initially, all the energy is stored as the mechanical energy of the impacting drop and surface energy of the sessile drop. Then, the mechanical energy of the system decreases, and is transferred into the surface energy of the drops. This transfer is followed by a recovery stage where surface energy is transferred back into the mechanical energy of the system. A part of the energy is lost as viscous dissipation. **(a)** Case I: $\chi = 0$, **(b)** Case II: $\chi = 0.08$, **(c)** Case III: $\chi = 0.25$, and **(d)** Case IV: $\chi = 0.625$. $E_m$ is the total mechanical energy of the system ($E_m = E_k + E_p$), $E_s$ the surface energy of the two drops and $E_d$ the viscous dissipation in the system. Note that the total mechanical energy ($E_m$) includes the energy of center of mass of the drops $\left(E_m^{CM} = E_k^{CM} + E_p\right)$ as well as the oscillation and rotational energies obtained in the reference frame that is translating with the center of mass of the individual drops.

These results indicate that the Direct Numerical Simulations provide a quantitative description of the impact dynamics. At this point, we investigate whether there is a one-to-one match of the experimental data and numerical simulations; this is done by comparing the drop boundaries and experimentally determined mechanical energies with the numerical predictions. Notably, we achieve a nearly quantitative agreement of the drop boundaries and experimental mechanical energies (Fig. 6).The different snapshots in Fig. 6 (i-iv) refer to the following time steps: (i) just at collision, (ii) sessile drop at maximum compression, (iii) droplet shape just before separation and (iv) final outcome of the impact. We expect that slight deviations between the experimental and numerically determined drop boundaries result from marginal inaccuracies in the experimental determination of the off-set parameter. However, the agreement is remarkably good, keeping in mind that there are no fitting parameters.

In Fig. 6a-v and 6b-v, we compare the measured experimental mechanical energies (data points) with those calculated using simulations (dotted lines). The calculated mechanical energies exceed the experimentally determined energies. To understand the origin of this discrepancy, one needs to consider that experimentally, we are only able to measure the vertical and horizontal displacements to approximate the mechanical energy of each drop. The images analysis did not offer an easy route to quantify the contribution of the rotational and oscillation energies that are included in the numerically calculated mechanical energy, $E_m$ . Therefore, to test whether neglecting the rotational and oscillation energies in our experiments causes the discrepancy, we calculated the center of mass mechanical energies ($E_m^{CM}$) for the two drops numerically (Fig. 6a-v and 6b-v, see *Materials and Methods* for a detailed discussion). The zero of the potential energy $\left(E_p^{CM} = 0\right)$ refers to the center of mass of the sessile drop at $t = 0$. This implies that $E_p^{CM}$ of the sessile drop becomes negative during compression. The center of mass kinetic energy ($E_k^{CM}$) is added to this value to get $E_m^{CM}$, namely, $E_m^{CM} = E_k^{CM} + E_p^{CM}$. As illustrated in Fig. 6a-v and 6b-v, the numerical results (solid lines) now nearly overlay the experimental results (data points). This holds for both the temporal development of the energy for the sessile drop as well of the impacting drop. We suppose that the small discrepancies may arise from finite adhesion of the sessile drop to the substrate (which is not accounted for in the simulations). An additional source of error may arise from the selection of time t = 0. We choose t = 0 based on the time instant when the sessile drop starts to feel the presence of the velocity field of the impacting drop, i.e., when the center of mass kinetic energy of the sessile drop becomes non-zero. Nevertheless, the remarkable agreement between the experimental and numerical results for the center of mass mechanical energies illustrate that the DNS are able to describe the oil drop-on-drop impact physics; this allows for quantifying the contribution of the rotational and oscillatory energies.

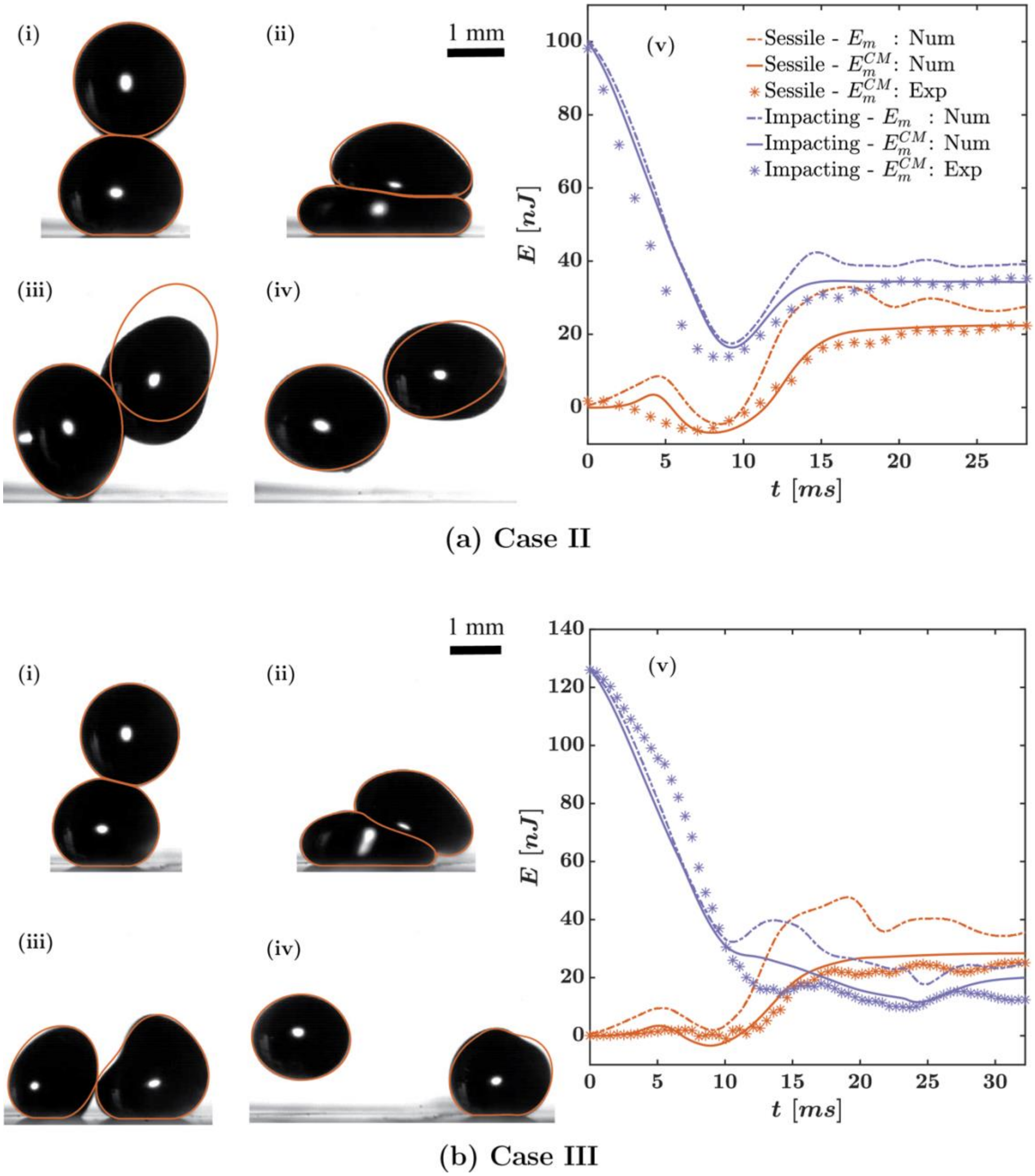

**Fig. 6. Validation of the numerical code: (a)** Case II: both sessile and impacting drop lift-off ($We \sim \mathcal{O}(1)$ & $\chi \approx 0.08$) for t = (i) 0 ms, (ii) 8 ms, (iii) 20 ms, (iv) 24 ms, and **(b)** Case III: sessile drop lifts-off and impacting drop rolls on the substrate ($We \sim \mathcal{O}(1)$ & $\chi \approx 0.25$) for t = (i) 0 ms, (ii) 8 ms, (iii) 20 ms, (iv) 24 ms. In the subfigures (i) to (iv), overlay of experimental images and DNS results (orange contour) are shown. (v) The mechanical energy of the center of mass ($E_m^{CM}$) calculated from experiments and simulations match within the experimental error**.** Note that in experiments, we could only keep track of the motion of the center of mass whereas in numerical simulations, the entire velocity field is known. Using this information, we can calculate the overall energy budgets. Here, the total mechanical energy of the drops ($E_m$) is shown in solid lines for reference. Error estimated in the experimental data is approximately 20% of the total energy.

## Conclusion:

By combining systematic experiments with numerical simulations, we illustrate how to predict and control the outcome of binary oil drop impacts on low adhesion surfaces. Four non-coalescing outcomes are attainable by varying the Weber number and the relative offset position of the impacting drops. One-to-one comparisons between the experimentally and numerically determined drop boundaries and center of mass mechanical energies illustrate the power of the Direct Numerical Simulations for quantitatively predicting the dynamics of drop-on-drop impact. More specifically, our numerical simulations illustrate that these general outcomes are governed by the average direction of the flow velocity vectors during the retraction phase, which are associated with the Weber number $We$ and the offset parameter $\chi$. In addition, our results illustrate that the ability to remove a sessile oil drop from the surface, as in Cases II and III, first requires sufficient energy transfer from the impacting drop and subsequently requires contrasting vector directions of the two retracting drops. Interestingly, our results illustrate that different outcomes exist even when the total dissipative losses of the system are similar. That is, the impact offset alone can be used to determine the recovered energy distribution between the two drops after impact.

## Materials and Methods

### Materials

The chemicals used are the following: ammonia (25% in water, Fluka), tetraethoxysilane (98%, Across Organics), trichloro(1H, 1H, 2H, 2Hperfluorooctyl) silane (97%, Sigma-Aldrich), acetone (Sigma-Aldrich), ethanol (> 99.8%, Sigma-Aldrich), toluene (Sigma-Aldrich), hexadecane (99%, Sigma-Aldrich). The chemicals were used as received. Milli-Q water was obtained from a Millipore purification system operating at 18.2 Mcm. Confocal microscope glass slides of $24 \times 60\ mm$ size and $170 \pm 5\mu m$ thickness were used (Carl Roth GmbH & Co.).

### Soot-templated superamphiphobic glass slides preparation

The soot-templated superamphiphobic glass slides were made following the process reported previously (*48*, *49*). The glass slides were sonicated for cleaning with ethanol, acetone, and toluene, for 5 min in each solvent. The glass slides were dried in an oven at 60ºC. For coating the glass slides with candle soot, the glass slides were hold above the center of the candle flame for approximately 1min. To form a uniform layer of soot particles, the glass slides were oscillatory moved in the horizontal plane. The coated glass slides were stored on a desiccator for 24 hrs. with an open snap cap vial containing 3mL of ammonium and a second vial with 3 mL of tetraethoxysilane. Afterwards, the samples were heated for 5hrs at 550ºC in an oven to get transparent substrates. The samples were coated with an approximately 25 nm thick silica shell. After activation in an oxygen plasma for 10min. the samples were fluorinated with trichloro (1H, 1H, 2H, 2Hperfluorooctyl) silane on a desiccator for 2 hrs.

### Scanning electron microscopy

Scanning electron microscopy (SEM) images were taken using a LEO 1530 Gemini and a SU800 Hitachi.

### Laser scanning confocal microscopy

Inverted laser scanning confocal microscopy (LSCM) images were taken with a Leica TCS SP8. The microscope was equipped with an HCX PL APO 40x/0.85 dry objective.

### Contact angle measurements

Roll-off angles measurements were performed using a goniometer OCA 35 for hexadecane drops of $5\mu L$. The apparent contact angle was measured with a Leica TCS SP8 confocal microscope for a hexadecane drop of $10\mu L$. The advancing and receding angles were measured while moving the hexadecane drop with a needle. The needle was supported on a micrometer stage next to the confocal microscope, as sketched in fig. S3. All angles were measured at least three times.

## Imaging and analysis of the impact

**Analysis for Fig. 2 and 3:** Subsequent images are recorded with two synchronized high-speed cameras to capture the evolution of the impact. The cameras were aligned perpendicular to each other to observe the impact in XZ and YZ planes. Both cameras, Photron Fastcam Mini UX100, were equipped with M Plan Apo (2×/0.055 ∞/0 f = 200) objective lenses. The cameras had a frame rate of either 2000 or 4000 fps, depending on the impact height, and at a resolution of 1280×1024 pixels. From the sequence of simultaneous side view images, the trajectory of the center of mass of each drop before impact was obtained. With this trajectory, the impact velocity $U_0$ and the separation distance $d$ were calculated. The tracking of the drops was done with an in-house developed MATLAB code. The image analysis started with a pre-processing step that involves contrast enhancement and noise removal in each frame. For noise removal, morphological closing and opening functions were applied to the grayscale images. The conversion of the grayscale images to binary images was done with a luminance threshold of 0.2. The analysis continues with drops detection, which included evaluation of the complement of the binary images, removal of objects with less than 5000 pixels, and filling holes. For drops tracking, the last step of the image analysis, MATLAB *regionprops* function was applied to find the coordinates of the centroids of each drop in all frames. Note that this function mixes the positions of the drops in subsequent frames, giving wrong trajectories. For obtaining the correct trajectories, the position for each drop in each frame was assigned such that the position satisfied the minimum distance between the centroids detected in subsequent frames.

**Analysis for Fig. 6a and 6b - (v):** Sequences of images are recorded with the aforementioned cameras. The experimental arrangement used is described in fig. S10. The setup, a modified version of setup in Fig. 1, allows a contrast between the drops during impact to delimit drop's interfaces. From the sequence of images, the trajectories of the drops were obtained. We use these trajectories to calculate the kinetic $E_k^{CM}$ and potential $E_p^{CM}$ energies. For tracking the drops, image pre-processing was done with the open source image analysis program FIJI. It starts with background subtraction and image inversion, with a threshold of 55%. Then, filling holes is applied. For drops detection, the *watershed* function is used to delimit drop's interfaces. With the *analyze particles* function, the drops are found in the images by setting the object size above 30 pixels. The coordinates of the drop's centroids are stock in separate files. This function mixes the positions of the drops in subsequent frames, displaying wrong trajectories. To correct drop tracking, we use the corresponding part of the aforementioned in-house developed MATLAB code.

## Simulation methodology:

We use a Finite Volume Method (FVM) based partial differential equation solver, Basilisk C (http://basilisk.fr/) for numerical simulation of incompressible Navier-Stokes equations (Eq. 3-4). All the equations are non-dimensionalized using the inertial-capillary velocity ($U_\gamma = \sqrt{\gamma/(\rho_l R)}$), radius of the impacting drop ($R$) and density of the liquid drops ($\rho_l$). Since we do not vary the type of liquid during and the volume of drops in our experiments or simulations, Ohnesorge number ($Oh = \mu_l/\sqrt{\rho_l \gamma R} = 0.0216$) and Bond number ($Bo =$ $\rho_l g R^2/\gamma = 0.308$) remain constant. Furthermore, in the simulations, the impact velocity is characterized by the impact weber number ($U_0 = \sqrt{We}$).

$$\frac{\partial U_i}{\partial X_i} = 0 \tag{3}$$

$$\frac{\partial U_i}{\partial t} + U_j \frac{\partial U_i}{\partial X_j} = \frac{1}{\hat{\rho}}\left(-\frac{\partial P}{\partial X_i} + Oh \frac{\partial\left(2\hat{\mu} D_{ij}\right)}{\partial X_j} + \kappa \delta_s n_i\right) + Bo\delta_{i3} \tag{4}$$

We use the geometric Volume of Fluid (VoF) (*50*) method for interface tracking. Consequently, one-fluid approximation (*51*) is used in the solution of the Navier-Stokes momentum equation (Eq. 4). In order to impose the condition of non-coalescence of the drops, different VoF tracers are used for the two droplets (Eq. 5, where $\{\Psi\} = \{\Psi_1, \Psi_2\}$. The use of two different tracers, along with interface reconstruction, ensures that there is always a thin air layer (thickness $\sim \Delta_1$, where $\Delta_1 = R/256$ is the size of smallest grid cell in the simulation domain). Similarly, in order to model the superamphiphobic substrate, it is assumed that there is a thin air layer (thickness $\sim \Delta_2$, where $\Delta_2 = R/512$ is the smallest grid cell near the substrate) between the drops and the substrate. All other boundaries are assumed to have no flow and free slip condition. We ensure convergence by comparing the viscous dissipation of the system and have chosen $\Delta$ such that the difference between consecutive simulations is small. The properties, such as density and viscosity are calculated using the VoF arithmetic property equations (Eq. 6, where $A_{gl}$ is the ratio of properties of gas and liquid).

$$\frac{\partial\{\Psi\}}{\partial t} + \frac{\partial(\{\Psi\}U_i)}{\partial X_i} = 0 \tag{5}$$

$$\hat{A}(\Psi_1, \Psi_2) = A_{gl} + (1 - A_{gl})(\Psi_1 + \Psi_2) \quad \forall A \in [\rho, \mu] \tag{6}$$

Basilisk C is a free software program. In this spirit, the authors would like to share the codes that have been used to simulate the cases reported in this manuscript. Detailed codes with documentation are available at https://github.com/VatsalSy/Lifting-a-sessile-drop. Please note that Basilisk C should be installed before running these codes.

**Energy calculations in the Direct Numerical Simulations:**

In this section, we discuss the different equations that we have used to calculate different energy budgets. First, we discuss the calculation of energies of the center of mass of the drops ($E_m^{CM}$),

$$E_m^{CM} = E_k^{CM} + E_p^{CM} \tag{7}$$

In Eq. 7, $E_k^{CM}$ and $E_p^{CM}$ are the center of mass kinetic energy and potential energy respectively. For these calculations, we first need to find the magnitude of velocity and position of the center of mass for each drop,

$$U_{CM} = \left|\frac{\int\int\int U_i d\,\Omega}{\int\int\int d\,\Omega}\right| \tag{8}$$

$$Z^{CM} = \frac{\int\int\int z d\Omega}{\int\int\int d\Omega} \tag{9}$$

In Eq. 8, the | | operator denotes the magnitude of the vector. In the above equations, $d\Omega$ is the differential fluid volume. Once $U_{CM}$ and $Z^{CM}$ are known, $E_k^{CM}$ and $E_p^{CM}$ can be calculated,

$$E_k^{CM} = \frac{2}{3}\pi U_{CM}^2 \tag{10}$$

$$E_p^{CM} = Bo\, Z^{CM} \tag{11}$$

The overall energy budget consists of the total mechanical energy $E_m = E_k + E_p$, the surface energy $E_s$, and the energy dissipation $E_d$, calculated as follows:

$$E_k = \iiint \left(\frac{1}{2}\hat{\rho}|U_i|^2\right) d\Omega \tag{12}$$

$$E_p = \iiint (\hat{\rho} Bo Z) d\Omega \tag{13}$$

$$E_s = \iint d\,\Gamma \tag{14}$$

$$E_d = \int_0^t \epsilon_\mu \; dt \tag{15}$$

In Eq. 12 and 13, energies of both the drops as well as the surrounding air medium are considered. Noticing that the density ratio of air to liquid, $\rho_{gl} = 1/770 \ll 1$ and that the domain is fixed in volume, the change in gravitational potential energy of the air medium is negligible. This implies that $E_p = E_p^{CM}$. In Eq. 14, $d\Gamma$ represents a differential surface. Lastly, Eq. 15 gives the total viscous dissipation in the system. In this equation, $\epsilon_\mu$ denotes the rate of dissipation at a given instant and is from

$$\epsilon_\mu = \iiint \left(2\hat{\mu} Oh |D_{ij}|^2\right) d\Omega \tag{16}$$

In the above equation, $|D_{ij}|$ is the second norm of the deformation tensor. The rate of viscous dissipation includes contributions from both the liquid drops and the air medium. In cases of drop impacts, the dissipation in air is important, especially in the thin air-layers between the drops, and between a drop and the substrate (3).

## Acknowledgments

D.V. & D.L. acknowledge the European Union's Horizon 2020 research and innovation program under the grant agreement No 722497, LubISS, J.T.P. the Alexander von Humboldt Foundation and University of Kentucky, and O.R.S. the Max Planck – Univ. Twente Center for Complex Fluid Dynamics for financial support and the Mexican National Council on Science and Technology (CONACyT) for the postgraduate fellowship. The authors acknowledge the ERC Advanced Grant No. 340391-SUPRO and 740479-DDD. The authors would like to thank Hans-Jürgen Butt for fruitful discussions. We would like to thank Michael Kappl for help with the estimation of forces on the substrate. We would also like to thank Abhishek Khadiya for support with measurements. V.S. would like to thank Andrea Prosperetti for insightful discussions on numerical simulations at different stages of the project, and Srinath Lakshman and Pierre Chantelot for discussion on energy dissipation in single drop impact phenomenon. Anke Kaltbeitzel, Gabriele Schäfer and Alexander Saal are acknowledged for technical support. The numerical simulations in this work were carried out on the Dutch national e-infrastructure with the support of SURF Cooperative.

## Authors' contributions

D.V. and J.T.P. planned the experiments and designed the structure of the manuscript. O.R.S. performed the experiments. D.L. and V.S. planned and performed the simulations. O.R.S. and V.S analyzed the data. J.T.P., D.V., V.S., O.R.S., and D.L. wrote the manuscript. The authors declare no financial conflicts of interest.

## Supplementary Materials

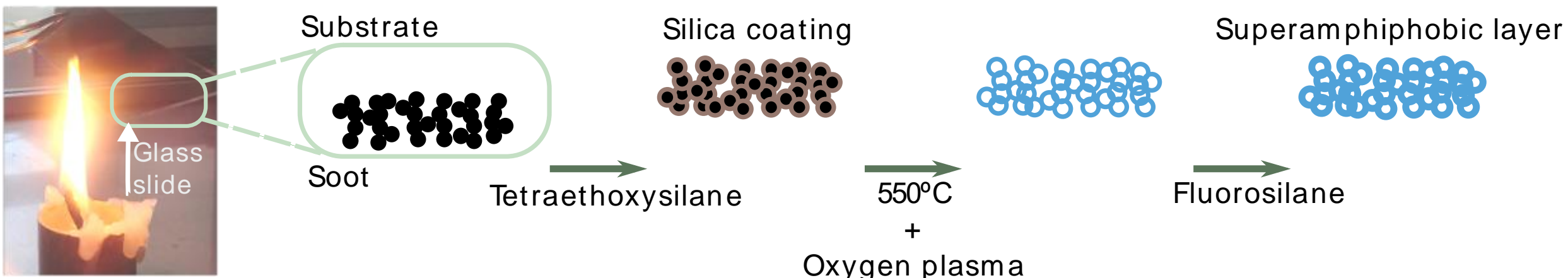

**Fig. S1. Preparation of soot-templated glass superamphiphobic surfaces:** Soot particles are deposited on a glass slide. The particles are coated with silica, by applying a chemical treatment with Tetraethoxysilane. To make the particle layer transparent, combustion is induced. With plasma treatment, OH groups are formed to chemically bind the trichloro (1*H,* 1*H,* 2*H,* 2*H*-perfluorooctyl) silane.

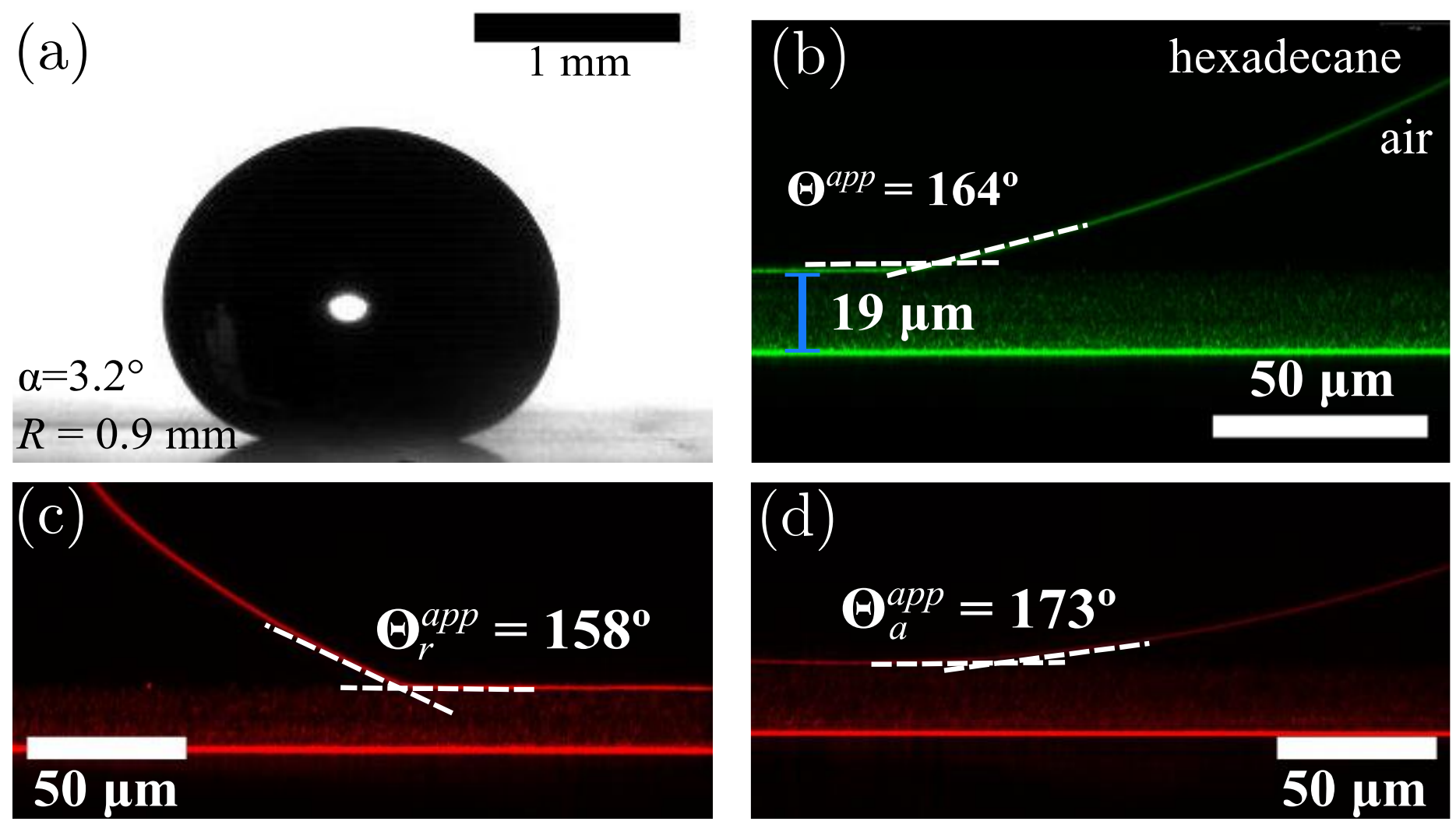

**Fig. S2. Sessile oil drop on a superamphiphobic substrate:** Shadowgraph **(a)** and confocal **(b-d)** images of a hexadecane drop on soot-templated glass slide. The shadowgraph image shows the typical shape of a sessile hexadecane drop during the experiments. The corresponding volumetric radius is 0.9 mm. The measured roll-off angle $\alpha$ of a drop of 5 μL is 3.2° (measured with a goniometer). The apparent contact angle $\Theta^{app} = 164°$, the receding angle $\Theta_r^{app} = 158°$, and the advancing angle $\Theta_a^{app} = 173°$, were measured with a drop of 10 μL size. Theoretically, $\Theta_a^{app}$ should be 180°. The observed difference results from the limited optical contrast.

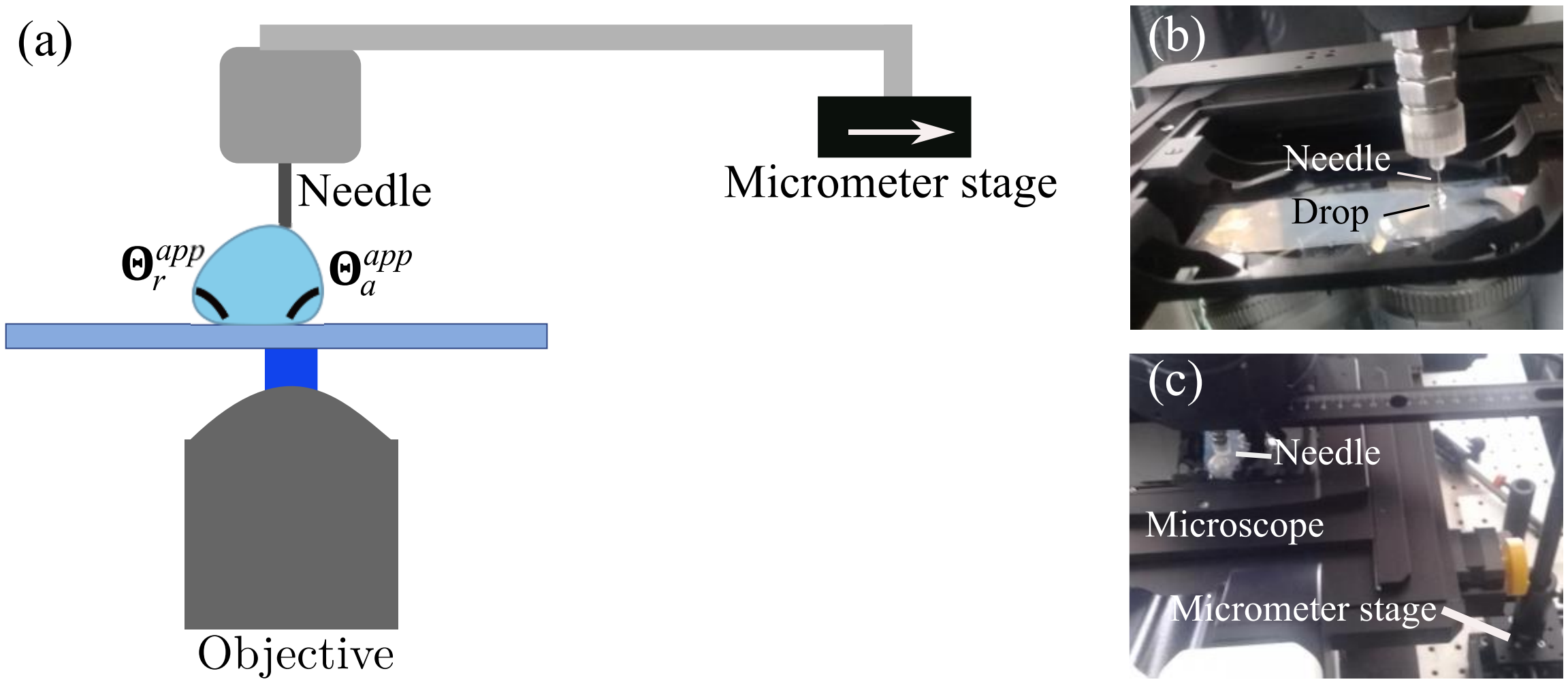

**Fig. S3. Sketch of the confocal microscope arrangement:** This arrangement is used to measure the apparent contact angle $\Theta^{app}$, the advancing $\Theta_a^{app}$ and receding $\Theta_r^{app}$ angles. A micrometer stage was adapted externally to the confocal microscope stage to sustain a needle from which the drops are hold and slowly dragged to observe $\Theta_a^{app}$ and $\Theta_r^{app}$.

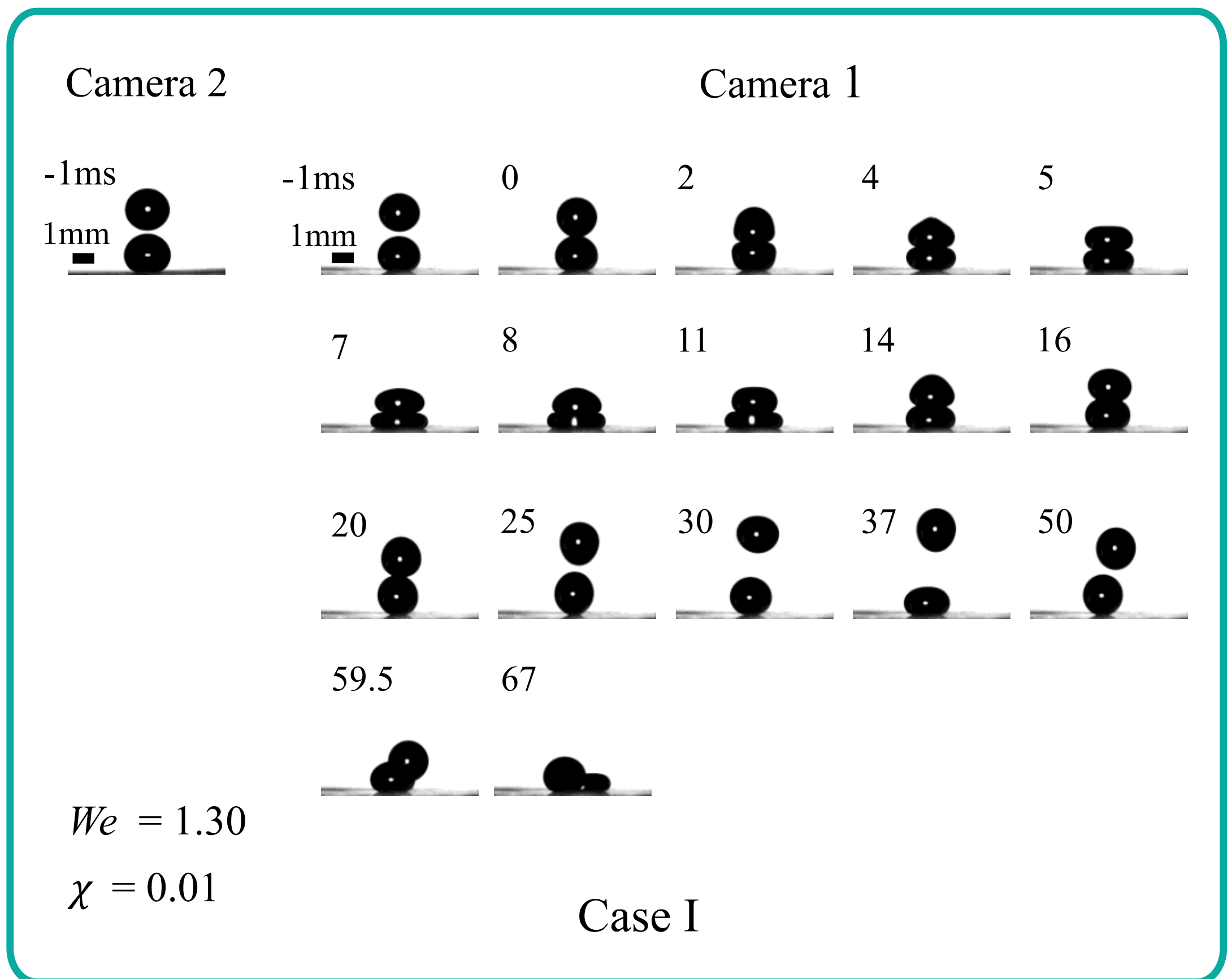

**Fig. S4. Time series of Case I for hexadecane drops:** pure bouncing of the impacting drop. The first row shows the compression of the drops during the spreading phase. The second row shows the retraction phase of both drops. In the third and fourth rows, the trajectory of the impacting drop is presented during and after bouncing. Snapshots of camera (1) and camera (2) at $t = -1$ ms show the alignment of the drops respect to each other.

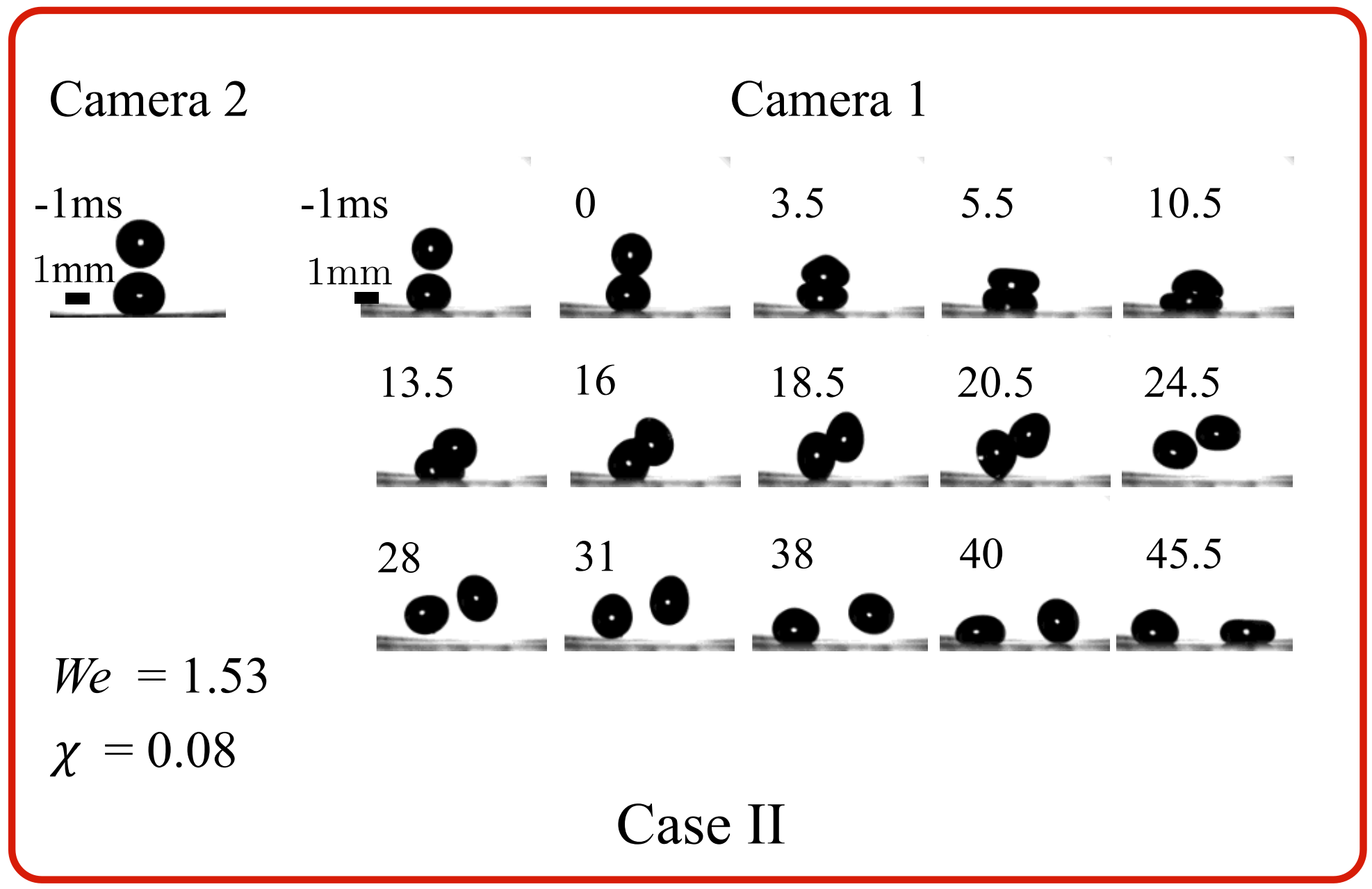

**Fig. S5. Time series of Case II for hexadecane drops:** Bouncing of the impacting drop followed by lifting-off of the sessile drop. The first row shows the compression of both drops during the spreading phase. The second row shows the retraction phase of both drops and at the end of the row the bouncing of the impacting drop is observed. In the third and fourth rows, the trajectory of the impacting drop is presented during and after bouncing as well as the trajectory followed by the sessile drop when lifting-off the substrate. Snapshots of camera (1) and camera (2) at $t = -1$ ms show the alignment of the drops.

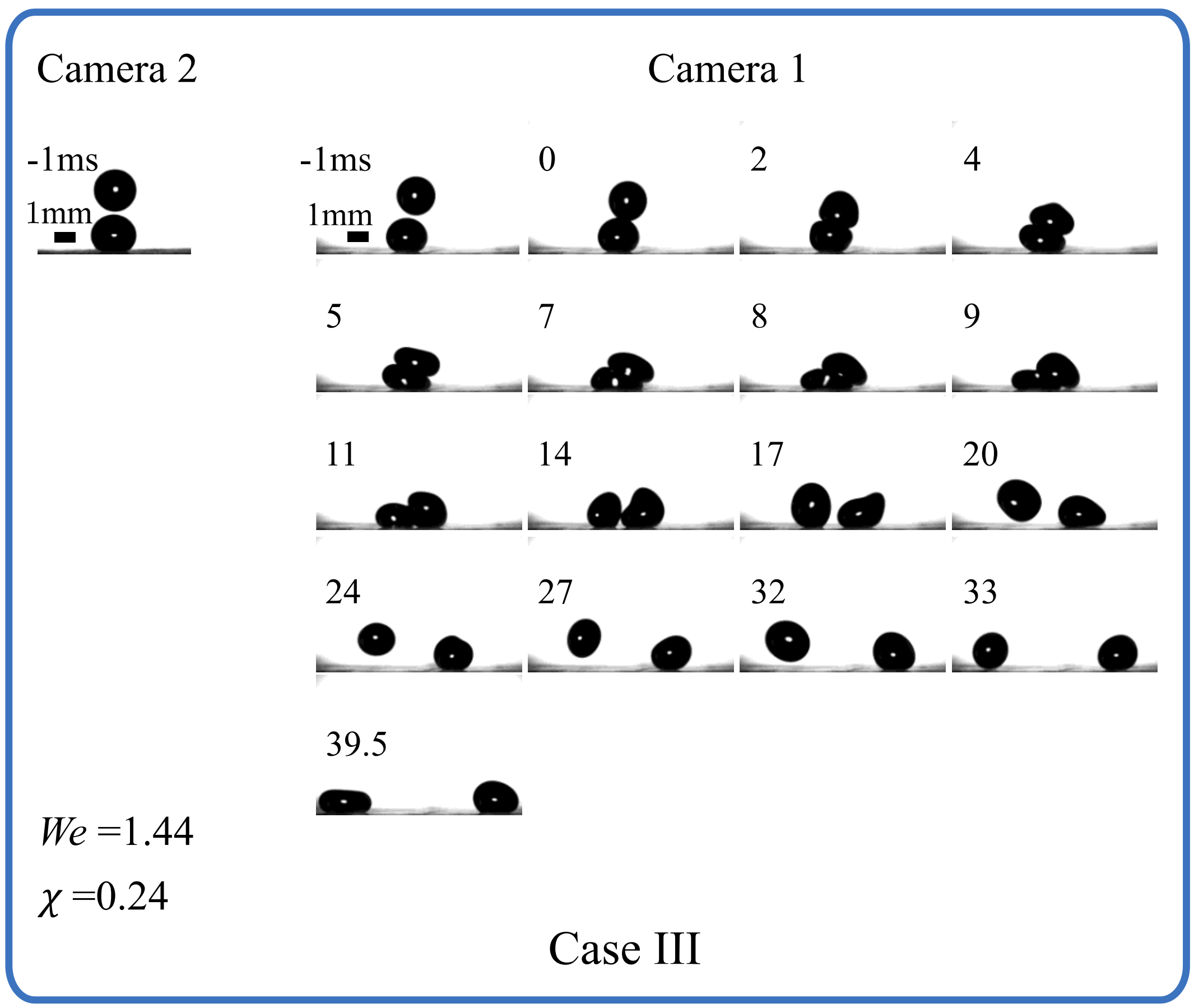

**Fig. S6. Time series of case III for hexadecane drops:** Rolling of the impacting drop on top of the sessile drop followed by lifting-off of the sessile drop. The first and second rows show the compression of the drops during the spreading phase. In the second row, the gliding of the impacting drop on the thin air-layer over the sessile drop is observed. In the third row, the rolling motion continues on the substrate while the sessile drop is in the retraction phase and lifts-off. In the fourth and fifth rows, the trajectory of the sessile drop during and after lifting-off is observed. Snapshots of camera (1) and camera (2) at $t$ = −1 ms show the alignment of the drops.

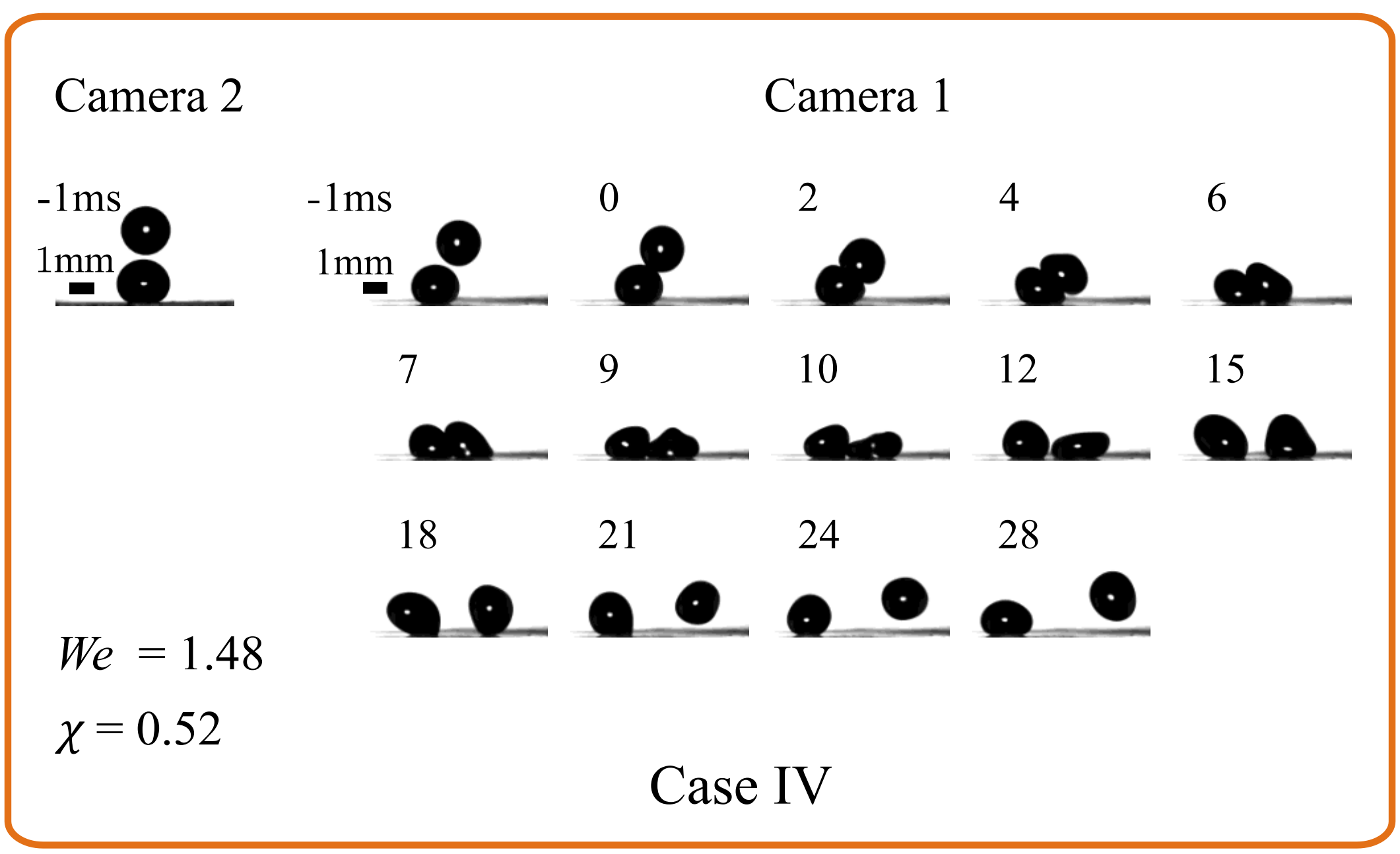

**Fig. S7. Time series of case IV for hexadecane drops:** Pure rolling of the impacting drop on top of the sessile drop. The first and second rows show the compression of the drops during the spreading phase. In the second row, the rolling of the impacting drop on top of the sessile drop and on the substrate is observed. In the second row, the retraction of the sessile drop is shown. In the fourth row, the displacement of the sessile drop in horizontal direction without jumping is observed while the impacting drop starts its retraction phase on the surface. The fourth row shows the rebound of the falling drop on the substrate after impacting with the sessile drop. Snapshots of camera (1) and camera (2) at $t$ = −1 ms show the alignment of the drops.

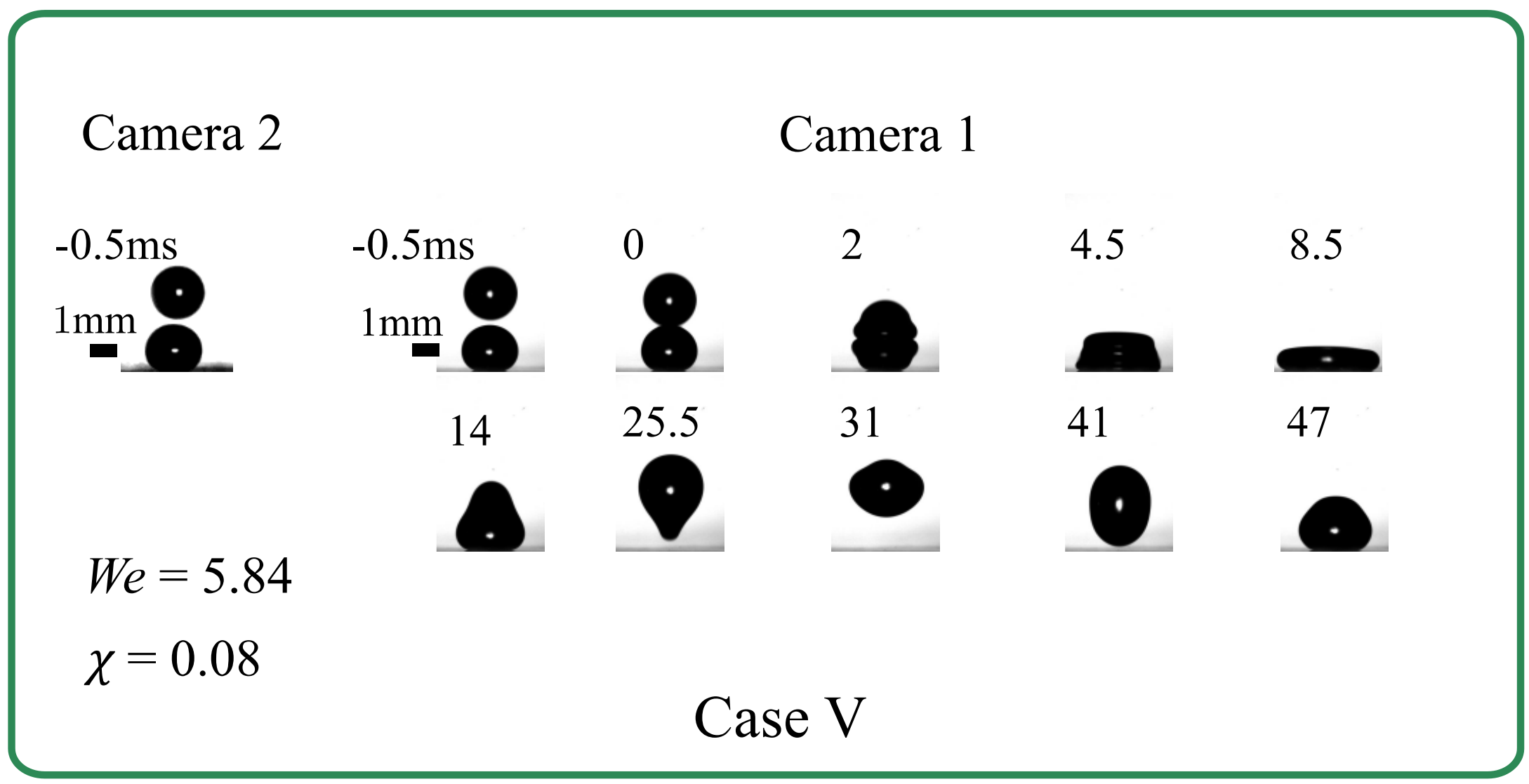

**Fig. S8. Time series of case V for hexadecane drops:** Coalescence between the impacting drop and the sessile drop followed by the detachment of the new drop from the substrate. The first row shows the compression of the drops during the spreading phase before coalescence ($t$ = 0 ms – 2 ms) and during coalescence ($t$ = 4.5 ms) until the maximum compression is reached ($t$ = 8.5 ms). The second row shows the retraction phase of the new drop and the detachment trajectory from the substrate. Snapshots of camera (1) and camera (2) at $t$ = −1 ms show the alignment of the drops.

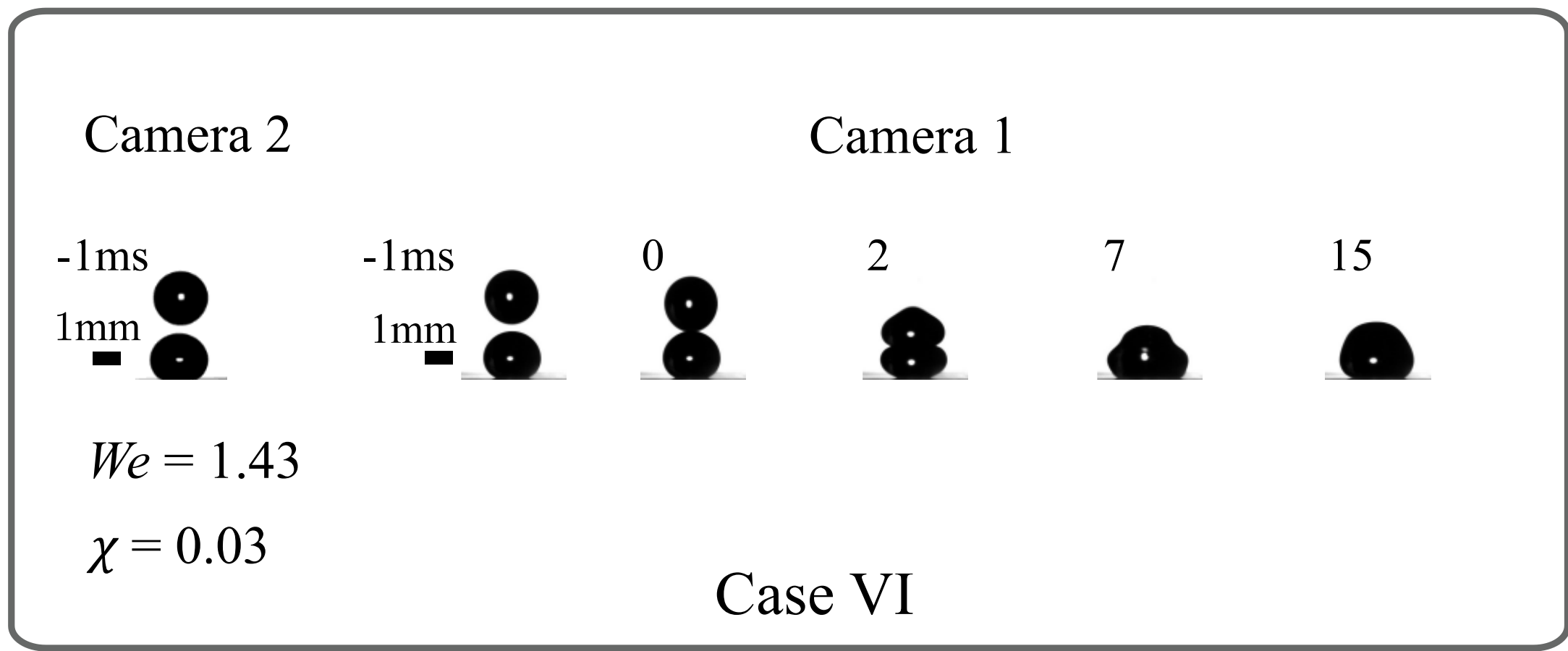

**Fig. S9. Time series of case VI for hexadecane drops:** Coalescence between the impacting drop and the sessile drop without detachment of the new drop from the substrate. The first three frames ($t$ = 0 ms – 2 ms) show the compression of the drops during the spreading phase before coalescence, the last two frames show during coalescence ($t$ = 7 ms – 15 ms). Snapshots of camera (1) and camera (2) at $t$ = −1 ms show the alignment of the drops.

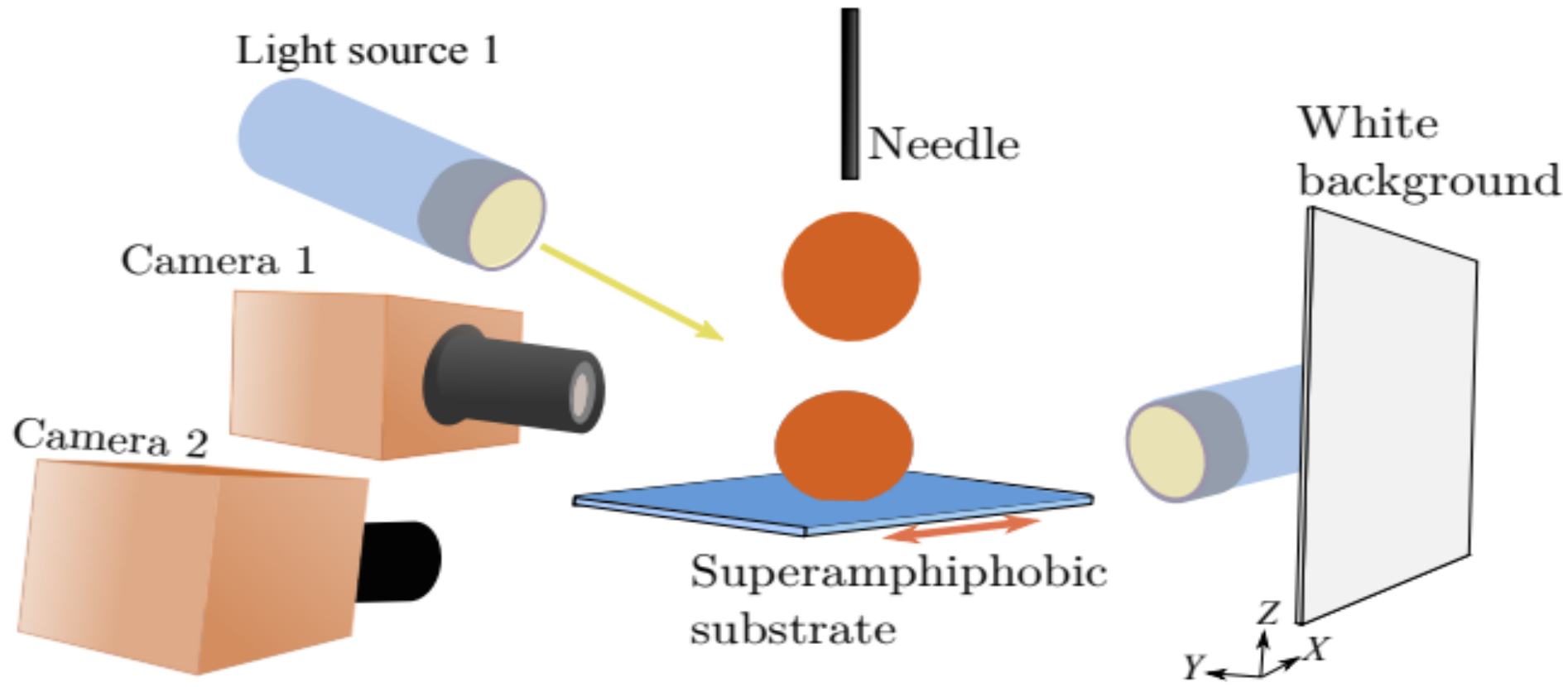

**Fig. S10.** Schematic of the experimental setup used to quantify the mechanical energy of the center of mass of the drops ($E_m^{CM}$). This setup is a modified version of the experimental setup shown in Fig. 1. Two high speed cameras are aligned perpendicular to each-other to record drop's trajectory in XZ and YZ planes. Camera 2 was used to center the impacting drop with the sessile drop in YZ plane. With Camera 1, the relative position was determined. Light source 1 was pointing at the same direction as Camera 1, both pointing to a white panel. The arrangement of Camera 1, light source 1 and the white panel allows a contrast between the drops during impact. With this contrast, the interface of the drops could be detected, hence the trajectory of the centroid of each drop. From these trajectories, the potential and kinetic energies were evaluated. Notice, the trajectories during impact are not possible to detect using the setup shown in Fig. 1, since both drops appear all black in the recorded images, hindering interface detection.

**Supplementary Movies:**

- **Movie 1:** (CaseI_Experiment.avi) Experimental video of Case I for hexadecane drops: bouncing of impacting drop. ($We \approx 1.30$ & $\chi \approx 0.01$)
- **Movie 2:** (CaseI_Numerics.mp4) Simulation video of Case I for hexadecane drops: bouncing of impacting drop. ($We = 1.50$ & $\chi = 0$)
- **Movie 3:** (CaseI_NumericsVelocityVectors.mp4) Simulation video showing velocity vectors of Case I for hexadecane drops: bouncing of impacting drop. The two-dimensional contour represents the slice $Y = 0$. Time is normalized by the capillary time scale, $t_\gamma = \sqrt{(\rho R_0^3)/\gamma}$. ($We = 1.50$ & $\chi = 0$)
- **Movie 4:** (CaseII_Experiment.avi) Experimental video of Case II for hexadecane drops: bouncing of the impacting drop followed by lift-off of the sessile drop. ($We \approx 1.53$ & $\chi \approx 0.08$)
- **Movie 5:** (CaseII_Numerics.mp4) Simulation video of Case II for hexadecane drops: bouncing of the impacting drop followed by lift-off of the sessile drop. ($We = 1.50$ & $\chi = 0.08$)
- **Movie 6:** (CaseII_NumericsVelocityVectors.mp4) Simulation video showing velocity vectors of Case II for hexadecane drops: bouncing of the impacting drop followed by lift-off of the sessile drop. The two-dimensional contour represents the slice $Y = 0$. Time is normalized by the capillary time scale, $t_\gamma = \sqrt{(\rho R_0^3)/\gamma}$. ($We = 1.50$ & $\chi = 0.08$)
- **Movie 7:** (CaseIII_Experiment.avi) Experimental video of Case III for hexadecane drops: sliding-off of the impacting drop on top of the sessile drop followed by lift-off of the sessile drop. ($We \approx 1.44$&$\chi \approx 0.24$)
- **Movie 8:** (CaseIII_Numerics.mp4) Simulation video of Case III for hexadecane drops: sliding-off of the impacting drop on top of the sessile drop followed by lift-off of the sessile drop. ($We = 1.50$ & $\chi = 0.25$)
- **Movie 9:** (CaseIII_NumericsVelocityVectors.mp4) Simulation video showing velocity vectors of Case III for hexadecane drops: sliding-off of the impacting drop on top of the sessile drop followed by lift-off of the sessile drop. The two-dimensional contour represents the slice $Y = 0$. Time is normalized by the capillary time scale, $t_\gamma = \sqrt{(\rho R_0^3)/\gamma}$. ($We = 1.50$ & $\chi = 0.25$)
- **Movie 10:** (CaseIV_Experiment.avi) Experimental video of Case IV for hexadecane drops: sliding-off of the impacting drop on top of the sessile drop followed by its lift-off. In this case, the sessile drop stays on the substrate. ($We \approx 1.48$ & $\chi \approx 0.52$)
- **Movie 11:** (CaseIV_Numerics.mp4) Simulation video of Case IV for hexadecane drops: sliding-off of the impacting drop on top of the sessile drop followed by its lift-off. In this case, the sessile drop stays on the substrate. ($We = 1.50$ & $\chi = 0.625$)
- **Movie 12:** (CaseIV_NumericsVelocityVectors.mp4) Simulation video showing velocity vectors of Case IV for hexadecane drops: sliding-off of the impacting drop on top of the sessile drop followed by its lift-off. In this case, the sessile drop stays on the substrate. The two-dimensional contour represents the slice $Y = 0$. Time is normalized by the capillary time scale, $t_\gamma = \sqrt{(\rho R_0^3)/\gamma}$. ($We = 1.50$ & $\chi = 0.625$)
- **Movie 13:** (CaseV_Experiment.avi) Experimental video of Case V for hexadecane drops: coalescence of drops and lift-off of coalesced drop. ($We \approx 5.84$ & $\chi \approx 0.08$)

- **Movie 14:** (CaseVI_Experiment.avi) Experimental video of Case VI for hexadecane drops: coalescence of drops and coalesced drop remains on the substrate. ($We \approx 1.43$ & $\chi \approx 0.03$)